\documentclass[bibyear]{aa} 

\usepackage{graphicx}
\usepackage{txfonts}
\def\mic{$\mu$m}

\begin{document}

   \title{Properties of extragalactic dust inferred from linear 
   polarimetry of Type Ia Supernovae\thanks   
   {Based on observations made with ESO Telescopes at the
   Paranal Observatory under Program IDs 076.D-0177 and 076.D-0178. This study is also based on observations collected at the German-Spanish  Astronomical 
   Center, Calar Alto (Spain) and on observations made with the Mercator 
   Telescope, Roque de los Muchachos, La Palma (Spain) equipped with the 
   HERMES spectrograph.}}

   \author{F. Patat\inst{1}
   \and 
   S. Taubenberger\inst{2}
   \and
    N.~L.~J. Cox\inst{3}
   \and
   D. Baade\inst{1}
   \and
   A. Clocchiatti\inst{4}
    \and
    P.~H\"oflich\inst{5}
    \and
    J.~R. Maund\inst{6}
    \and
    E. Reilly\inst{6}
    \and
    J. Spyromilio\inst{1}
    \and
    L. Wang\inst{7}
    \and
    J.~C. Wheeler\inst{8}
    \and
    P. Zelaya\inst{4}
}

   \offprints{F. Patat}

   \institute{European Organisation for Astronomical Research in the 
        Southern Hemisphere (ESO), Karl-Schwarzschild-Str. 2,
              85748 Garching b. M\"unchen, Germany\\
              \email{fpatat@eso.org}
              \and
              Max Planck Institute for Astrophysics (MPA), 
              Karl-Schwarzschild-Str. 1, 85748 Garching b. M\"unchen, Germany
              \and
              Instituut voor Sterrenkunde, KU Leuven, Celestijnenlaan 200D, 
              bus 2401, 3001 Leuven, Belgium
              \and
              Departamento de Astronomia y Astrofisica, PUC,
              Casilla 306, Santiago 22, Chile 
              \and
              Department of Physics, Florida State University, Tallahassee,
              Florida 32306-4350, USA             
              \and
              Queen's University Belfast, Belfast, BT7 1NN, UK
              \and
              Department of Physics, Texas A\&M University, College Station, 
              Texas 77843, USA
              \and
              Department of Astronomy and McDonald Observatory, 
              The University of Texas at Austin, Austin, TX 78712, USA
             }

   \date{Received April, 2014; accepted February, 2014}

\abstract
{}
{The aim of this paper is twofold: 1) to investigate the properties of
  extragalactic dust and compare them to what is seen in the Galaxy;
   2) to address in an independent way the problem of the anomalous
  extinction curves reported for reddened Type Ia Supernovae (SN) in
  connection to the environments in which they explode.}
{The properties of the dust are derived from the wavelength dependence
  of the continuum polarization observed in four reddened Type Ia SN:
  1986G, 2006X, 2008fp, and 2014J. The method is based on the observed
  fact that Type Ia SN have a negligible intrinsic continuum
  polarization. This and their large luminosity makes them ideal tools
  to probe the dust properties in extragalactic environments.}
{All four objects are characterized by exceptionally low total-to-selective absorption ratios ($R_V$)
  and display an anomalous interstellar polarization law, characterized by
  very blue polarization peaks.  In all cases the polarization
  position angle is well aligned with the local spiral structure. While
  SN~1986G is compatible with the most extreme cases of interstellar polarization known in the
  Galaxy, SN~2006X, 2008fp, and 2014J show unprecedented
  behaviours. The observed deviations do not appear to be connected to
  selection effects related to the relatively large amounts of
  reddening characterizing the objects in the sample.}
{The dust responsible for the polarization  of these four SN is most likely of
  interstellar nature. The polarization properties can be interpreted
  in terms of a significantly enhanced abundance of small grains. The
  anomalous behaviour is apparently associated with the properties of the
  galactic environment in which the SN explode, rather than with the
  progenitor system from which they originate. For the extreme case of SN~2014J,
  we cannot exclude the contribution of light scattered by local material; however,
  the observed polarization properties require an ad hoc geometrical dust distribution.}


\authorrunning{F. Patat et al.}
\titlerunning{Properties of extragalactic dust inferred from linear polarimetry}

   \maketitle
%

\begin{table*}
\centering
\tabcolsep 1.7mm
\caption{\label{tab:sample}Basic properties of the SN sample}
\begin{tabular}{llllcccccccc}
\hline
SN  & Host & Host & Host & $R_V$    & $A_V$  & $E_{B-V}$&Ref. & $P_{04}$   & $\lambda_{max}$  & $P_{04}/E_{B-V}$ & $\theta$\\
    &      & type & notes&   &        &         &     & (\%)      &  ($\mu$m)        & (\% mag$^{-1}$) & (deg)\\
\hline
1986G  & NGC5128     & S0, pec &starburst &2.57$^{+0.23}_{-0.21}$ & 2.03$^{+0.09}_{-0.13}$ &0.79$^{+0.08}_{-0.08}$ &1&  5.1$\pm$0.1 & 0.43   & 6.5$\pm$0.7 & 117.9$\pm$0.8\\
2006X  & NGC4321     & SAB(s)bc& &1.31$^{+0.08}_{-0.10}$ & 1.88$^{+0.09}_{-0.13}$ &1.44$^{+0.11}_{-0.15}$ &1&  7.8$\pm$0.2 & $<$0.4 & 5.4$\pm$0.5 & 139.5$\pm$0.1\\
2008fp & ESO 428-G14 & SAB(r)0 &active &1.20$^{+0.26}_{-0.14}$ & 0.71$^{+0.10}_{-0.08}$ &0.59$^{+0.15}_{-0.10}$ &1&  2.6$\pm$0.1 & $<$0.4 & 4.4$\pm$1.0 & 148.4$\pm$0.2\\
2014J  & NGC3034     & I0      & starburst& 1.40$\pm$0.10         &  1.85$\pm$0.11         &1.37$\pm$0.03          &2&  6.6$\pm$0.1 & $<$0.4  & 4.8$\pm$0.1 & 42.2$\pm$0.3\\
\hline\multicolumn{12}{l}{References: (1) Phillips et al. \cite{phillips13}; (2) Amanullah et al. 
\cite{amanullah}.}
\end{tabular}
\end{table*}

\section{\label{sec:intro}Introduction}

Studying cosmic dust has important consequences on our understanding
of a number of astrophysical processes, ranging from galaxy evolution
to stellar and planetary formation. The characterization of dust in
the diffuse interstellar medium relies heavily on the observed
wavelength dependencies of extinction and polarization (see
Voshchinnikov \cite{vosh12}, for a comprehensive review). While this
study can be undertaken in our Galaxy using pair-matching of stars of
the same spectral and luminosity classes (see Whittet \cite{whittet03},
for a review), its application to the extragalactic case is only
possible for very few nearby systems, like the Magellanic clouds. At
larger distances, other methods are used: foreground extinction of
distant quasars, differential extinction of multiply lensed quasars,
gamma-ray-burst/supernova host extinction, and extinction in star-burst
galaxies.

In this article, we focus on the use of Type Ia Supernovae (SN). These
bright ($M_V\sim -$19) objects were used several times as
line-of-sight probes to infer the properties of the intervening
material. The technique applied so far can be considered as an
extension of the pair-matching method, in which the observed
spectrophotometric properties are compared to a set of un-reddened
templates. This has led to the striking result that the
total-to-selective absorption ratio ($R_V$) is systematically and
significantly lower than what is typical in the Galaxy ($R_V\sim$3;
Fitzpatrick \& Massa \cite{fitz}). From the analysis of 80 Type Ia SN
with $E_{B-V}\leq$0.7, Nobili \& Goobar (\cite{nobili}) derived an
average value $R_V$=1.75$\pm$0.27. Although this result was revised
towards more ``normal'' values ($R_V=$2.8$\pm$0.3, Chotard et
al. \cite{chotard}), objects with relatively large reddening
($E_{B-V}>$0.4) show unusually low $R_V$ (Phillips et
al. \cite{phillips13}). This has interesting implications both for our
understanding of dust and the explosion environments of Type Ia SN.

Here we discuss an independent approach, which relies on the fact that
SN Ia have a very low intrinsic continuum polarization ($<$0.3\%,
Wang \& Wheeler \cite{WW08}). This level of continuum polarization is
negligible when the reddening is significant ($E_{B-V}>$0.1-0.2) and the 
interstellar polarization prevails. Their luminosity, insignificant intrinsic
continuum polarization, and the supposed absence of significant amounts of
circumstellar material (at variance with core-collapse events) with
possibly peculiar characteristics, make Type Ia SN suitable background
illuminators to probe the dust properties in extragalactic
environments.

The paper is structured as follows. In Section~\ref{sec:obs} we
present the sample. Section~\ref{sec:isp} discusses the continuum
polarization dependence and compares it to a sample of Galactic stars.
The significance of the observed deviations from the Galactic behaviour
and their physical implications are analysed in
Section~\ref{sec:disc}. Finally, our conclusions are summarized in
Section~\ref{sec:conc}.

\section{\label{sec:obs}Data}

The number of reddened Type Ia SN ($E_{B-V}>$0.5) with polarimetric
data of sufficient quality (in terms of accuracy and wavelength
coverage) is very small. To the best of our knowledge this includes
only SN 1986G (NGC~5128; Cen A), 2006X (NGC~4321; M100), 2008fp
(ESO~428-G14), and the recent 2014J (NGC~3034; M82). Other Type Ia SN
have spectropolarimetric data: 1997dt, 2002bf, 2003du, 2004dt
(Leonard et al. \cite{leonard05}); 2001el (Wang et al. \cite{wang03});
2007sr (Zelaya et al. \cite{zelaya13}); 2005ke (Patat et
al. \cite{patat12}); 2011fe (Smith et al. \cite{smith12}); 2012fr
(Maund et al. \cite{maund13}); however, the low reddening associated
with these events makes the study of the interstellar polarization
(ISP) wavelength dependence very uncertain, if not impossible (see
also Zelaya et al. \cite{zelaya14}).

The four selected objects show a pronounced continuum polarization at
position angles remarkably well aligned with the local spiral arms of their
hosts, as expected for dust grain alignment along the galactic
magnetic field (Scarrot, Ward-Thompson \& Warren-Smith
\cite{scarrot}). In all cases, the wavelength dependence of the
polarization position angle is very mild or null (see
Figure~\ref{fig:angle}), similar to what is seen in Galactic stars
(Dolan \& Tapia \cite{dolan}). All four SN Ia have high-quality,
high-resolution spectroscopy data, with a number of separate
interstellar absorption components indicating significant amounts of
gas (atomic and molecular) at different velocities along the
line-of-sight (D'Odorico et al.  \cite{sandro}; Cox \& Patat
\cite{cox08}; Cox \& Patat \cite{cox14a}; Cox et al. \cite{cox14b};
Goobar et al. \cite{goobar14}; Welty et al. \cite{welty14}).

The relevant properties of the sample discussed in this paper are
summarized in Table \ref{tab:sample}. In addition to the reddening
characteristics, the table reports the basic polarization parameters,
which will be introduced and discussed in Section~\ref{sec:isp}.

\subsection{SN 1986G, 2006X, and 2008fp}

SN~1986G is the first Type Ia SN that was used for studying the
polarization properties of the intervening dust. For this object,
UBVRIJH broad-band polarimetry was obtained by Hough et
al. (\cite{hough87}). SN 2006X and 2008fp were observed in the
framework of our spectropolarimetric programme with the FOcal
Reducer/low-dispersion Spectrograph (FORS1), mounted at the Cassegrain
focus of the ESO--Kueyen 8.2m telescope (Appenzeller et al.
\cite{appenzeller}). The data and their reduction are discussed in
Patat et al. (\cite{patat09a}) and Cox \& Patat (\cite{cox14a}).

\subsection{SN~2014J}

Spectropolarimetry of SN~2014J is reported here for the first time. A
preliminary analysis was published by Patat et al. (\cite{patat14}).
A full analysis of the spectropolarimetric data will be presented in
a separate paper, while here we focus purely on the continuum
polarization.

We obtained linear spectropolarimetry of SN~2014J  on three epochs
(Jan 28, Feb 03, and Mar 08, 2014), using the Calar Alto Faint Object
Spectrograph (CAFOS), mounted at the 2.2 m telescope in Calar Alto,
Spain (Meisenheimer \cite{cafos}; Patat \& Taubenberger
\cite{patat11}, hereafter PT11). We obtained all spectra  with the
low-resolution B200 grism coupled with a 1.5 arcsec slit, giving a
spectral range 3300-8900 \AA, a dispersion of $\sim$4.7 \AA\/
px$^{-1}$, and a full width half maximum (FWHM) resolution of 21.0 \AA. For each retarder angle,
we obtained two exposures; integration times ranged from 10 to 25
minutes per exposure. Because of the low signal-to-noise ratio and the
known instrumental limitations (PT11), the CAFOS spectropolarimetric
data below 3800 \AA\/ cannot be used.

\begin{figure}
\centering
\includegraphics[width=9cm]{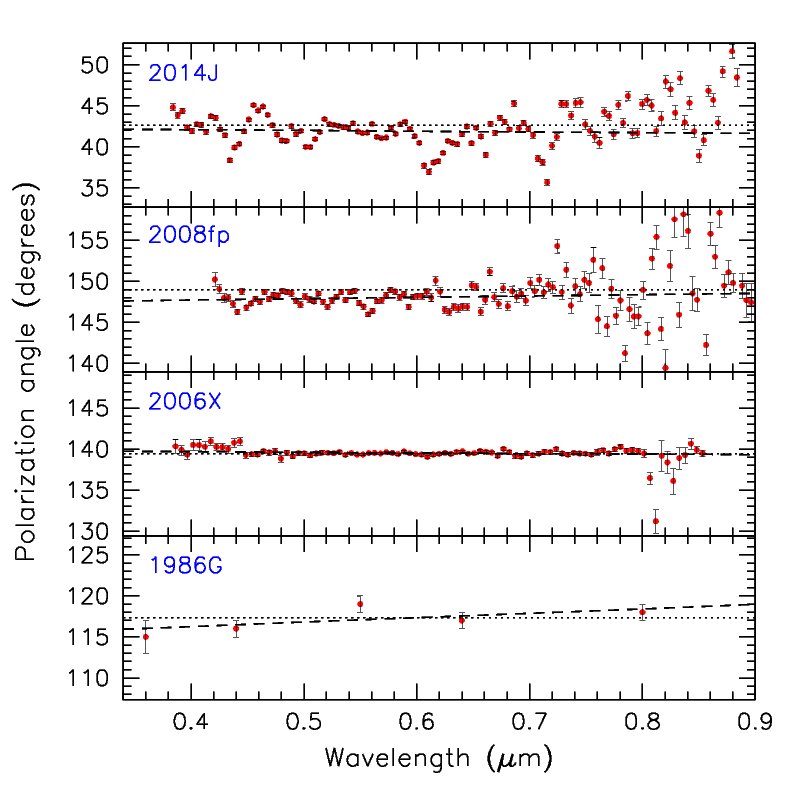}
\caption{\label{fig:angle}Polarization angles for the SNe in the
  sample. The dashed lines trace the weighted linear best fit
  ($\lambda\leq$0.8 \mic), while the dotted lines indicate the average
  value computed using the full wavelength range.}
\end{figure}

The Stokes parameters were derived with the same procedure as used for
SN 2006X and 2008fp. We peformed error estimates  following the
prescriptions described by Patat \& Romaniello (\cite{patat06}), while
the HWP zero-point angle chromatism was corrected using tabulated data
(PT11). To check and correct for instrumental polarization, we observed the
unpolarized standard HD~90508 (Serkowski \cite{serk1974}) 
at each epoch. We detected spurious polarization of 0.3\%. Although
the polarization level of HD~90508 is constant ($\sim$0.25\%; see also PT11),
different corrections were applied to the three data sets, as the
polarization angle changed between the epochs. For this reason, the
correction is accurate to a few 0.1\%. Given the high level of
polarization displayed by SN~2014J (see below), this does not affect
any of the conclusions that we present. The overall
performance of the system and the quality of the chromatic and
instrumental polarization corrections were checked using observations
of the polarized standard star HD~43384 (Mathewson \& Ford
\cite{mathewson}). The maximum polarization is
$P_{max}$=3.10$\pm$0.05\% at $\lambda_{max}$=0.539 \mic\/ and the average position
angle $\theta$=170.3$\pm$0.2 degrees. These values are fully
consistent with those reported by Serkowski, Mathewson,
\& Ford (\cite{serkowski}): $P_{max}$=3.0\%,
$\lambda_{max}$=0.53 \mic, $\theta$=170 degrees. The marked wavelength dependence of the
position angle ($d\theta/d\lambda$=7.4$\pm$0.6 degrees $\mu$m$^{-1}$)
is consistent with the findings of Doland \& Tapia (\cite{dolan}).

For our purposes, we use the data of the second epoch
that, being obtained closest to maximum light, have the best
signal-to-noise ratio. The first and the last epoch show  very
similar behaviour, with the changes related to the evolution of the
intrinsic SN properties, most prominently in the \ion{Si}{ii} 6355 \AA\/
and the \ion{Ca}{ii} near-IR triplet. The continuum polarization level is
constant within the measurement errors.

The signal is dominated by a continuum polarization that grows steadily and
rapidly from the red to the blue, reaching about 6.6\% at 0.4
\mic. The average polarization position angle for SN~2014J is
42.2$\pm$0.3 degrees, which is well aligned with the local spiral
structure and in good agreement with the value reported by Greaves et
al. (\cite{greaves}) for the dust lane (40 degrees). In particular, it
is fully consistent with the position angle distribution along the
dust lane of M82 (Jones \cite{jones00}; see his Figures 4 and 5) and
identified as being generated by transmission (as opposed to
scattering).

\section{\label{sec:isp}ISP wavelength dependence}

\begin{figure*}
\centering
\includegraphics[width=16cm]{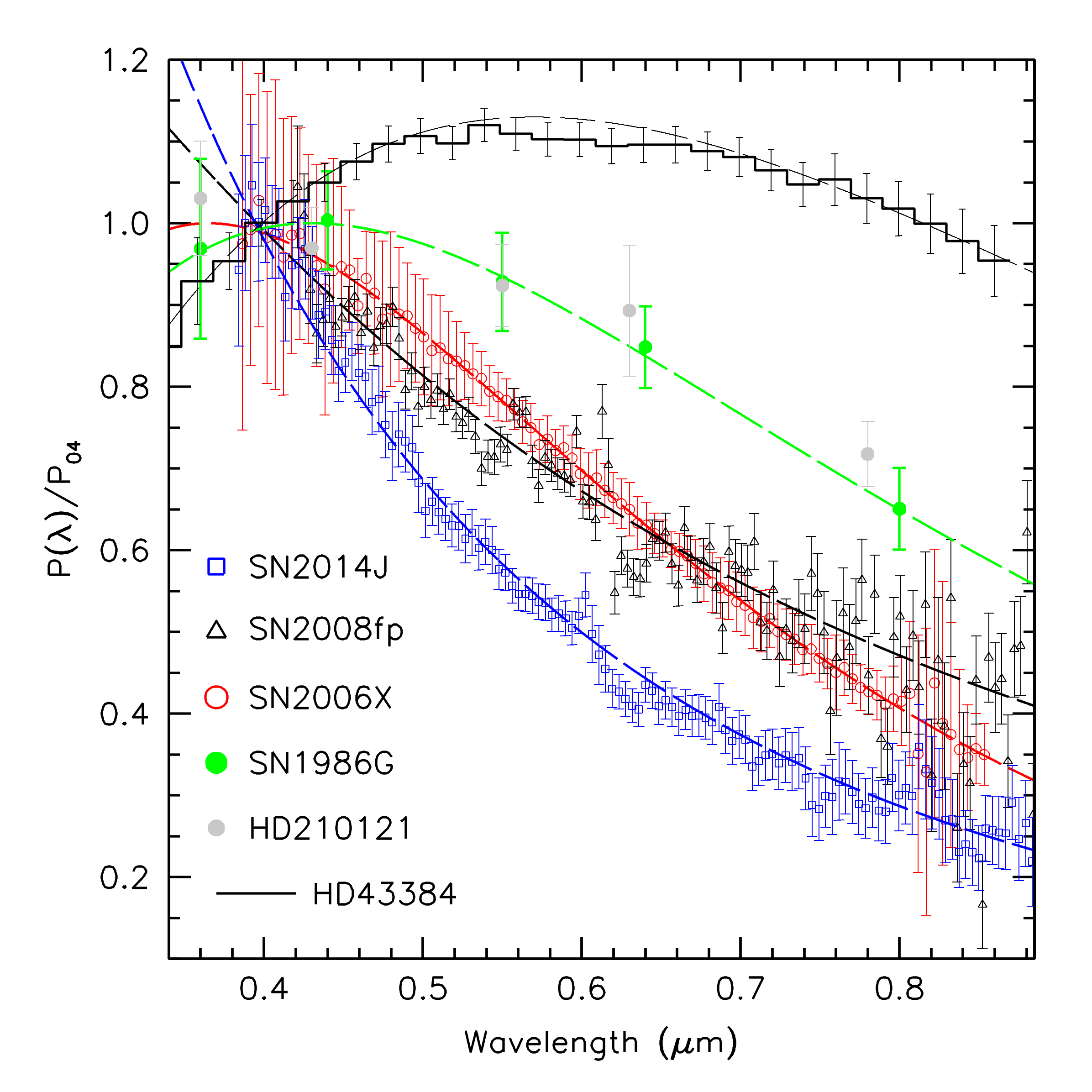}
\caption{\label{fig:isp} Observed polarization wavelength
  dependence for SN~1986G (Hough et al. \cite{hough87}), SN~2006X
  (Patat et al. \cite{patat09a}), SN~2008fp (Cox \& Patat
  \cite{cox14a}), and SN~2014J (this work).  For comparison, the data of
  HD~43384 (this work) and HD~210121 (Larson et al. \cite{larson96})
  are also plotted. The dashed curves trace Serkowski law best
  fits. For presentation, the curves were normalized to $P_{04}$ (see
  text). The error bars were kept at their original values. HD~43384 is a 
  polarized standard; HD~210121 is a Galactic star with very low $R_V$.
  For SNe 2006X, 2008fp and 2014J, $P(\lambda)$ was computed as the
  polarization component parallel to the polarization angle.}
\end{figure*}

In Figure~\ref{fig:isp} we present the wavelength dependence displayed
by the four SN. The deviation from the typical behaviour seen in the
Milky Way (MW) is illustrated by the comparison with the Galactic star HD~43384
(solid histogram). While SN~1986G shows a polarization maximum at
$\lambda_{max}$=0.435 \mic, SN 2006X, 2008fp, and 2014J do not show
evidence of a peak in the spectral range covered by the data. In the
absence of a maximum, we quantify the polarization level with the
value measured at 0.4 \mic\/ ($P_{04}$). This is then used to compute
the polarization efficiency $P_{04}/E_{B-V}$, which ranges from 4.4 to
6.5 \% mag$^{-1}$ (see Table~\ref{tab:sample}). Since the polarization
maximum is not observed, these are only lower limits for the true
polarization efficiency. It is therefore not possible to firmly
establish whether or not the SN conform to the Galactic limit
$P_{max}/E_{B-V}<$9 \% mag$^{-1}$ (Serkowski et
al. \cite{serkowski}). The weighted average $P_{04}/E_{B-V}$ is
4.9$\pm$0.4 \% mag$^{-1}$. Given the relatively large values of
$E_{B-V}$, this estimate is robust in contrast to low-extinction
cases (see the discussion in Leonard et al. \cite{leonard02}).  For
presentation, the polarization curves are normalized to $P_{04}$. The
errorbars were kept at their original amplitudes to allow for a
direct comparison of the different uncertainties characterizing the
various data sets.

The exceptional behaviour displayed by Type Ia SN compared Galactic stars is made
more evident by considering the
wavelength dependence of the most extreme known cases.  To the best of
our knowledge, HD~193682 has the bluest ISP peak reported in the
literature, with $\lambda_{max}\leq$0.32 \mic\/ (Anderson et
al. \cite{anderson}); however, a more recent study revised this
finding, re-positioning the maximum at 0.445$\pm$0.020 \mic\/
(Weitenbeck \cite{weitenbeck08}). Other interesting cases are those of
Cyg~OB2 No.~10 and 12, for which Whittet et al. (\cite{whittet})
report $\lambda_{max}$=0.33 \mic\/ and 0.35 \mic, respectively. The
polarization wavelength dependence seen in these two objects resembles
that of SN~1986G, possibly with a larger slope (see also the best fit
presented by Martin et al. \cite{martin92} for Cyg~OB2 No.~12:
$\lambda_{max}$=0.35$\pm$0.02 \mic, $K$=0.61$\pm$0.04).  Another
relevant case is that of HD~210121, for which independent estimates of
polarization and total-to-selective extinction are available. This
star was reported to have $\lambda_{max}$=0.38$\pm$0.02 \mic\/ and an
exceptionally low $R_V$=2.1$\pm$0.2 (Larson, Whittet \& Hough
\cite{larson96}). In general, the wavelength dependence is again
similar to that of SN~1986G. In the optical and near-UV, the two data
sets are consistent within the one-sigma errors (see
Figure~\ref{fig:isp}, gray filled circles), while the near-IR data
indicate a slightly flatter curve for HD~210121.

As already discussed in Patat et al. (\cite{patat09a}), SN~2006X shows
a steadily growing polarization to the blue, reaching about 8\% at the blue edge
of the spectral coverage. The curve flattening observed in the blue
suggests that the maximum polarization occurs shortly before 0.38
\mic. A non-linear least-squares fit using the Serkowski empirical
expression ($P(\lambda)/P_{max}=\exp[-K
  \ln^2(\lambda_{max}/\lambda)]$) gives $K$=1.47$\pm$0.05 and
$\lambda_{max}$=0.365$\pm$0.02 \mic. Although the derived value for
$K$ is completely different from that predicted by the Whittet et
al. (\cite{whittet}) relation (0.62$\pm$0.07), the resulting curve gives an
excellent match to the data (see Figure~\ref{fig:isp}).  

The case of SN~2008fp is similar to that of SN~2006X (see also Cox \&
Patat \cite{cox14a}); however, the nature of the data and the
increased noise in the red (partially due to a marked fringing) do not
allow a well-constrained fitting with a Serkowski law (the formal
solution is rather degenerate, with very elongated contours in the
$\chi^2$ space). We note that probably one third of the reddening
suffered by this object arises in the Milky Way (see also Cox \& Patat
\cite{cox14a}), so that the red portion of the curve may be affected
by the Galactic component. The slope observed in the blue favours a
value of $K$ smaller than that of 2006X, with a possible maximum below
0.4 \mic. A formal least-squares fit using the Serkowski formulation
yields $K$=0.40$\pm$0.05 and $\lambda_{max}$=0.148$\pm$0.03 \mic.

Finally, SN~2014J appears to be the most extreme case, with a very
steep polarization increase in the blue. With the lack of UV data, it is
impossible to extrapolate the position of the polarization peak.  The
Serkowski law was shown to approximately hold in the UV for Galactic
stars with $\lambda_{max}$ in the optical and UV coverage (see
Anderson et al. \cite{anderson}; Martin, Clayton \& Wolff
\cite{martin99}).  However, with only the possible exception of
HD~193682, there are no published direct observations of Galactic 
objects with peaks in the UV. The best fit for SN~2014J has $K$=0.40$\pm$0.07 and
$\lambda_{max}$=0.05$\pm$0.02 \mic. For both SN 2008fp
and 2014J, the derivation of the Serkowski parameters is based on the
unjustified assumption that the expression can be extrapolated to
extremely short wavelengths. The best-fit curves are plotted in
Figure~\ref{fig:isp} for pure illustrative purposes (see next
section).

In summary, with the possible exception of SN~1986G, the
wavelength dependence seen in our SN sample has never been observed in
the Galaxy. A similar conclusion was reached by Kawabata et al. (\cite{kawabata}),
based on the published data for 1986G, 2006X and 2008fp and broad-band polarimetry
for 2014J.

\begin{figure*}
\centering \includegraphics[width=16cm]{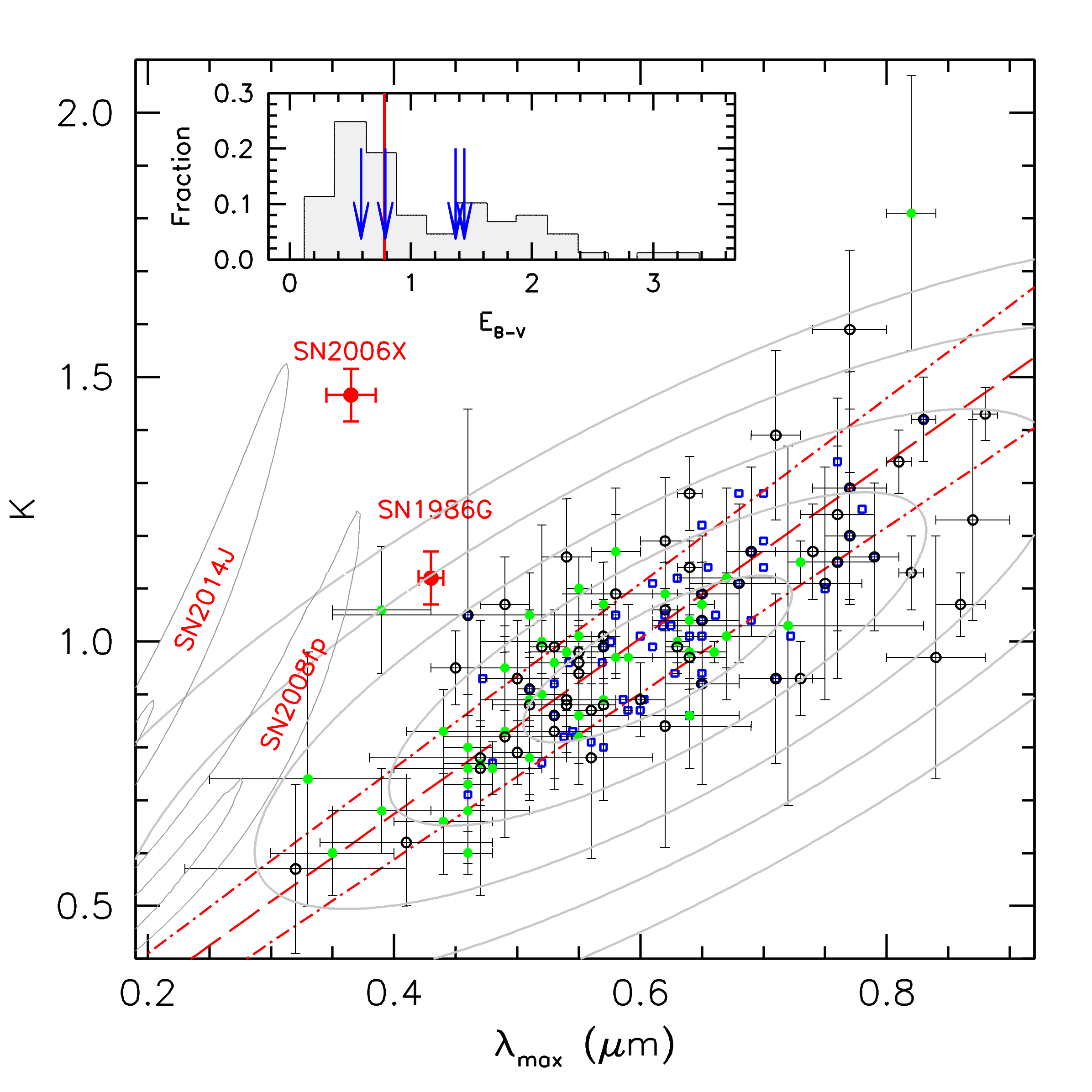}
\caption{\label{fig:kappa} Serkowski parameters $\lambda_{max}$ and
  $K$ for Galactic stars. The data are from Whittet et
  al. (\cite{whittet}, circles) and Voshchinnikov et al. (\cite{vosh13},
  squares).  Filled symbols indicate objects with $E_{B-V}>$0.8. The ellipses trace 
  the 1- to 5-sigma confidence levels for
  the full Whittet et al. sample (for a bi-variate Gaussian distribution). For comparison, the parameters
  derived for SN 1986G and 2006X are plotted (large filled
  circles). The long-dashed line traces the Whittet et al. relation
  (the dashed-dotted lines indicate the $\pm$1-sigma levels). The
  contours in the lower left corner of the plot indicate the 10-
  and 20-sigma confidence levels for SN 2008fp and 2014J. The upper
  left insert plots the $E_{B-V}$ distribution of the Whittet et
  al. (\cite{whittet}) sample of 105 stars. The vertical line marks the median
  value of $E_{B-V}$, while the arrows indicate the values of the four
  SN (see Table~\ref{tab:sample}).}
\end{figure*}

\section{\label{sec:disc}Discussion}

\subsection{Statistical significance}

Although we have shown with some confidence that all four SN in our sample deviate from
the average behaviour seen in the Galaxy, it is important to quantify
the significance of this deviation.  For this purpose, in
Figure~\ref{fig:kappa} we present the Galactic samples by Whittet et
al.  (\cite{whittet}) and Voshchinnikov et al. (\cite{vosh13}) on the
$\lambda_{max}-K$ plane. The second sample has some intersection with
the first and no error bars are given. It is therefore included only
for the sake of completeness and it is not used in the statistical
analysis that follows.

As concluded by Whittet et al. (\cite{whittet}; see also Wilking,
Lebofsky, \& Rieke \cite{wilking}), the Galactic data cluster around a linear
relation (K=0.01$\pm$0.05 + [1.66$\pm$0.09]$\lambda_{max}$). This
empirical law (see Figure~\ref{fig:kappa}, long-dashed line) was
derived by Whittet et al. (\cite{whittet}) rejecting 27 objects from
the original sample that includes\ 105 stars. For quantifying the
bi-dimensional distribution, we have estimated the probability
contours from the error-weighted covariance matrix computed on the
full Whittet et al. sample. These are traced as light-colored ellipses
at 1-sigma intervals. The misalignment between the best-fit Whittet et
al. relation and the major axis of the ellipses is due to the
inclusion of the 27 objects rejected by Whittet et
al. (\cite{whittet}).

Out of the total sample, 1.9\% of the Galactic objects fall outside of the
4-sigma level. Of the five stars with $\lambda_{max}<$0.4 \mic, four have
$K$ values that are consistent with the derived relation. The other,
Cyg OB~2 A, is off by 4 sigma ($\lambda_{max}$=0.39$\pm$0.04 \mic,
$K$=1.06$\pm$0.12; $E_{B-V}$=2.59). Interestingly, for this star
Whittet et al. (\cite{whittet}) report that it ``has a polarization
curve rising anomalously steeply to the blue, and lies furthest from
the best straight line''. Although the wavelength dependence is very
similar to that shown by SN~1986G, for this star there are no U-band
data and the B point is characterized by a relatively large
uncertainty (P=6.6$\pm$0.7\%). At face value, there are no signs of a
maximum in the optical domain for this star.

Cy OB 2 A represents slightly less than 1\% of the Whittet et
al. (\cite{whittet}) sample. This fraction decreases to $\sim$0.7\% if
one includes the Voshchinnikov et al. (\cite{vosh13}) sample, giving a
first quantitative estimate of the rarity of SN~1986G-like
polarizations in our Galaxy. As first pointed out by Hough et
al. (\cite{hough87}), this indicates that the dust along the line of
sight to SN~1986G is significantly different from what has been
studied and published so far for the Milky Way. SN~2006X constitutes 
an even more deviant case (Patat et
al. \cite{patat09a}). The polarization law for this SN appears to be at
more than 9 sigma from the MW sample (see Figure~\ref{fig:kappa}).

The cases of 2008fp (Cox \& Patat \cite{cox14a}) and 2014J are more
difficult to place on the $\lambda_{max}-K$ plane, because of the
degeneracy of the Serkowski law fitting caused by very weakly
constrained peak positions (see previous section). Nevertheless, one
can get an indication by running a brute-force $\chi^2$ mapping as a
function of $\lambda_{max}$ and $K$. The resulting 10- to 20-sigma
confidence levels are plotted in Figure~\ref{fig:kappa} (left lower
corner). As anticipated, they are very elongated, hence implying a
wide range of suitable parameter combinations. Although this is
strictly valid only for a Serkowski law (and there is no guarantee
that this holds when the maximum is in this extremely blue regime),
the contour plots indicate that both objects deviate in a very
significant way from the Galactic behaviour. Given the similarities
between SN 2006X and 2008fp (see Figure~\ref{fig:isp}), we argue that
SN~2008fp deviates from the Galactic sample at a similar statistical
significance level.

This analysis confirms the findings presented in the previous section:
all four SN in our sample display exceptional properties. If they were all
drawn from the Galactic distribution, then having four outlier events
at the observed significance levels would be very unlikely. At face
value this leads to the conclusion that they belong to a different
underlying population.

\subsection{Selection effects}

\begin{figure}
\centering \includegraphics[width=9cm]{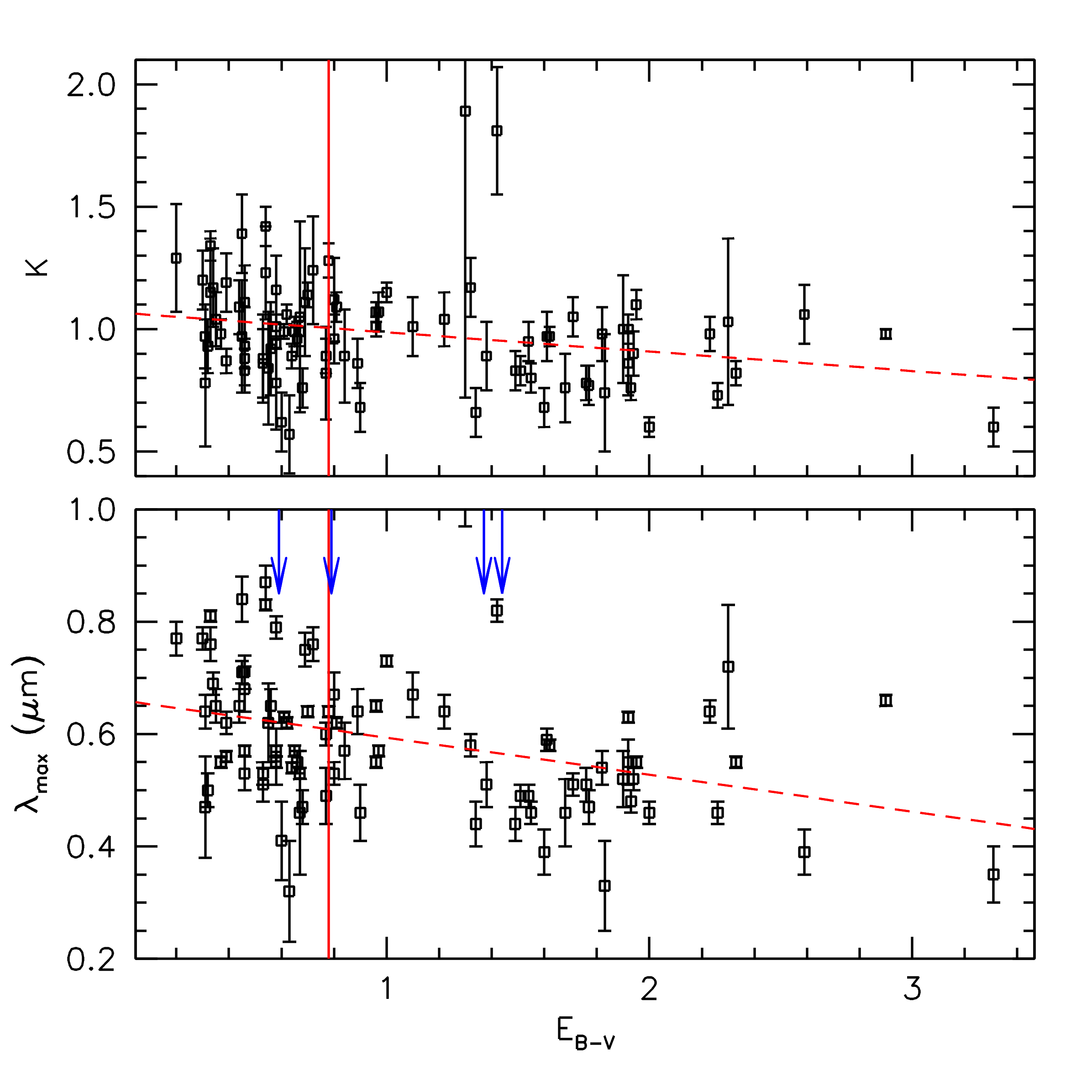}
\caption{\label{fig:ebv}Serkowski parameters $K$ and $\lambda_{max}$
  as a function of $E_{B-V}$ for the Whittet et al. (\cite{whittet})
  sample. The dashed lines trace the best linear least-squares fit to
  the data. The vertical solid line marks the median $E_{B-V}$, while
  the arrows are placed at the values of the four SN (see
  Table~\ref{tab:sample}).}
\end{figure}

The current sample of SN is small and possible selection effects may be
playing an important role. Because of the way the sample was constructed,
the four SN discussed in
this paper are all significantly reddened (0.6$\lesssim
E_{B-V}\lesssim$1.4; see Table~\ref{tab:sample}). One may therefore
argue that this introduces systematic biases with respect to the
Galactic sample, hence favouring special $\lambda_{max},K$
combinations.  The $E_{B-V}$ distribution for the Whittet et
al. sample is shown in the insert of Figure~\ref{fig:kappa} and has a
median $E_{B-V}\simeq$0.8. The stars with a reddening larger than the
median tend to show smaller values of $\lambda_{max}$ and $K$ (see the
filled circles in Figure~\ref{fig:kappa}). However, the linear
correlation coefficients of $\lambda_{max}$ and $K$ vs.The  $E_{B-V}$ are
-0.34 and -0.24, respectively, indicating a weak dependence of these
parameters on the reddening (see Figure~\ref{fig:ebv}). In addition,
$\sim$30\% of the Whittet et al. stars have $E_{B-V}>$1.4, i.e. larger
than the highest value in the SN sample (see
Table~\ref{tab:sample}). These facts imply that, at least for the
Galactic sample, selecting highly reddened objects does not lead to
strong systematic effects on the final determination of the $K$
vs. $\lambda_{max}$ relation, or on the overall distribution of the
selected events on the $\lambda_{max}-K$ plane.

We note that the ISP wavelength dependence observed in several
supernovae, not only of Type Ia, clearly deviates from the Galactic
Serkowski law (Leonard \& Filippenko \cite{leonard01}; Maund et
al. \cite{maund07a}; Patat et al. \cite{patat09b}; Maund et
al. \cite{maund10}). Notably, the Type II SN~1991gi shows a {\em
  \textup{normal}} wavelength dependence ($K$=1.15 and $\lambda_{max}$=0.51
\mic, Leonard et al. \cite{leonard02}). This SN suffered  a minimal
reddening ($E_{B-V}$=0.21$\pm$0.09, with a robust upper limit of
$E_{B-V}<$0.45). Although in the light of the above findings this may
be more an exception than the rule, the low reddening experienced by
this object may again indicate that the strong deviations occur in
highly obscured events.

Another possible source of bias resides in the host
characteristics. Both NGC~5128 (S0, peculiar) and NGC~3034 (I0, edge
on) are starburst galaxies, a fact that may have some relevance for
the dust properties (see for instance Hutton et
al. \cite{hutton14}). While SN~2014J in NGC~3034 is the most discordant object,
SN~1986G in NGC~5128 is the least deviant. The host of SN~2006X is a normal
SAB(s)bc spiral, while 2008fp exploded in a peculiar SAB(r)0 galaxy,
with radio jets and signs of activity (Veron \& Veron \cite{veron}).
Given the exiguous size of the sample, it is impossible to tell
whether there is a systematic correlation between the host properties
and the ISP behaviour.

Finally, one can question whether the observed behaviours are due to the
peculiar nature of isolated clouds along the lines-of-sight. In both
SN 2008fp and 2006X (that incidentally show a similar ISP wavelength
dependence), the bulk of reddening is produced within molecular clouds
signaled by exceptionally strong CN features (Cox \& Patat
\cite{cox08,cox14a}; see also the discussion in Phillips et
al. \cite{phillips13}). Molecular gas was also detected towards
SN~1986G in the form of weak CH and CH$^+$ absorptions, although no CN
features were reported (D'Odorico et al. \cite{sandro}). The situation
is very similar for SN~2014J, in which CN is very weak (Welty et
al. \cite{welty14}; Sternberg et al. \cite{assaf}).

\begin{figure*}
\centering \includegraphics[width=16cm]{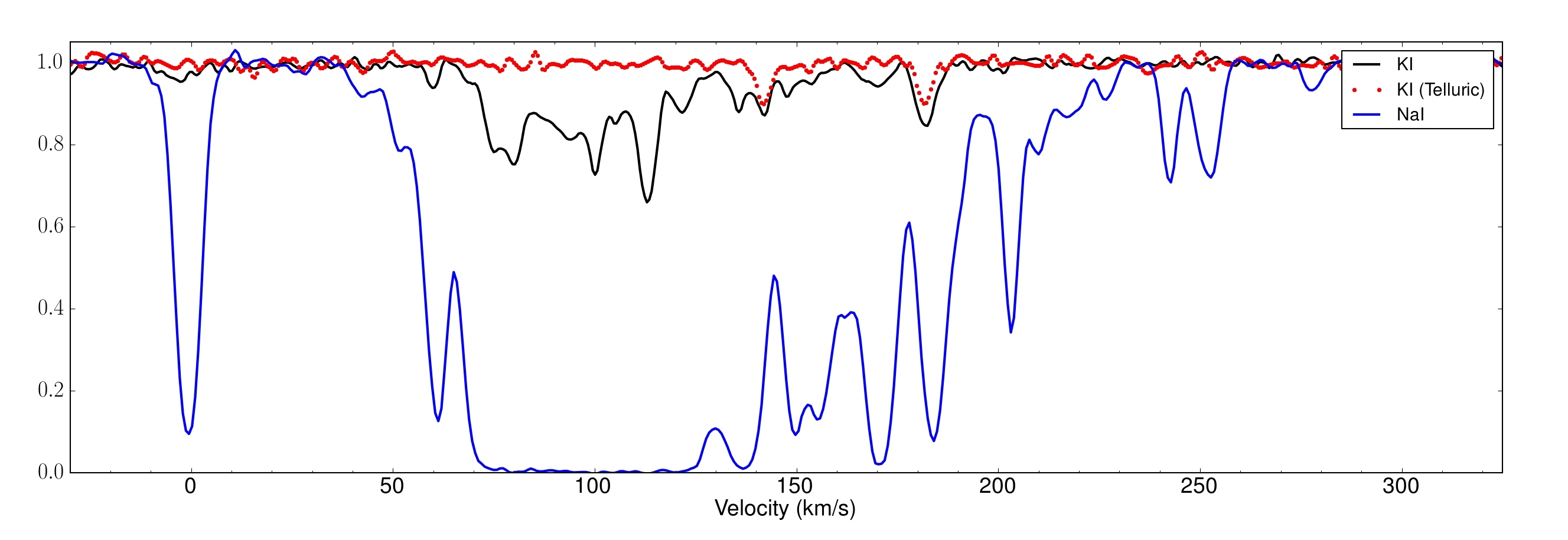}
\caption{\label{fig:sn2014J} Galactic and interstellar absorption
  components for Na~ID (blue) and K~I (black). The dotted curve traces
  a telluric absorption spectrum for the K~I region. The data were
  obtained with the HERMES spectrograph (Sternberg et al. \cite{assaf}.)}
\end{figure*}

In the case of 2006X and 2008fp, the bulk of reddening
(and hence polarization) most probably originates within one single
large thick cloud, which is not the case for SN 1986G and 2014J. For
these two objects high-resolution spectroscopy reveals a number of
\ion{Na}{i} and \ion{Ca}{ii} features with comparable depth. This is
illustrated for 2014J in Figure~\ref{fig:sn2014J}, which presents the
absorption profiles of Na and K lines (see also Welty et
al. \cite{welty14}). In addition to the Galactic component at zero
velocity, the profile shows more than ten extragalactic
absorptions. Even within the strongly saturated \ion{Na}{i}~D feature,
the weaker \ion{K}{i} line reveals more than five distinct components
of comparable strength. This excludes that the bulk of reddening (and
polarization) is generated within one single thick cloud with peculiar
properties. The observed behaviour results from the combined presence
of numerous clouds along the line-of-sight, spanning a wide range of
velocities (and most probably positions) within the host. For this
reason, we conclude that what is observed in SN~2014J reflects the
average properties of the disk of M82, which is seen almost
edge-on. This conclusion probably also applies  to SN~1986G, which displays
a similarly large number of components (D'Odorico et al. \cite{sandro}).

\subsection{Implications on dust properties and extinction law}

Before addressing the implications of the findings presented in the
previous sections, it is important to note that deviations from the
empirical Serkowski law can be obtained by combining the effects of
two or more clouds with different properties (see Coyne \& Gehrels
\cite{coyne}; see also Patat et al. \cite{patat09b}; Appendix B). For
instance, one can imagine a case in which the observed polarization
results from the combination of two dust systems, individually
described by different combinations of $P_{max}$, $\lambda_{max}$, $K$
and position angle $\theta$. Depending on the values of these
parameters, the output polarization may show very deviant wavelength
dependencies. Normally this is accompanied by a wavelength dependence
of the position angle (like the one observed for HD~43384; see
Section~\ref{sec:obs}). Hough et al. (\cite{hough87}) reported a
  mild dependence ($d\theta/d\lambda$=4.5$\pm$1.9 degrees \mic$^{-1}$)
  for 1986G, which may signal that the polarization position angles of
  the various clouds along the line of sight are different. For the
  other objects in the sample, $d\theta/d\lambda$ is consistent with
  zero (2006X: -0.8$\pm$0.5 degrees \mic$^{-1}$; 2008fp: +1.7$\pm$1.4;
  2014J: -0.9$\pm$1.6 degrees \mic$^{-1}$ ; see also
  Figure~\ref{fig:angle}). It is important to note that there is no
way any combination of Serkowski laws with $\lambda_{max}$ in the
wavelength interval found in our Galaxy can produce polarization
curves like those observed for SN 2006X, 2008fp, and 2014J. This
implies that there must be different reasons for the observed behaviour.

The observed linear dependence of $K$ on $\lambda_{max}$ is
qualitatively understood as a systematic increase in the number of
small, aligned grains in regions displaying bluer polarization peaks
(Martin et al. \cite{martin99}; Whittet et al. \cite{whittet01};
Voshchinnikov \cite{vosh12}). This argument has been used, for
instance, to infer that the size of dust grains along the
line of sight to SN~1986G is $\sim$20\% smaller than what is typical
in the Milky Way (Hough et al. \cite{hough87}). This is more
quantitatively confirmed by recent modelling. Siebenmorgen,
Voshchinnikov, \& Bagnulo (\cite{ralf}; their Figure 5) show that
substantial changes in the polarization law take place in the optical
and UV domain when the minimum grain size decreases from 200 nm to 50
nm, with the polarization peak moving from $\sim$1 \mic\/ to 0.25
\mic, respectively. Analogous results are presented by
  Voshchinnikov \& Hirashita (\cite{vosh14}; their Figure 2). 
  The application of this line of reasoning to the cases of SN 2006X, 2008fp, and 2014J, 
  for which  $\lambda_{max}<$0.4 \mic, leads to a dust size reduction
 $>$ 27\%. However, the modelling also demonstrates that dust size is
not the only ingredient (see below).

It has been suggested for a long time that the shape of the polarization 
law is related to the extinction law. Serkowski et
al. (\cite{serkowski}) found that the total-to-selective absorption
ratio relates to the position of the polarization peak, following the
empirical relation $R_V=5.5\lambda_{max}$. This was slightly revised
by Whittet \& van Breda (\cite{whittet78};
$R_V$=[5.6$\pm$0.3]$\lambda_{max}$) and Clayton \& Mathis
(\cite{clayton}; $R_V$=$-$0.29$\pm$0.74 +
[6.67$\pm$1.17]$\lambda_{max}$). Inserting $\lambda_{max}\leq$0.4
\mic\/ into these empirical relations gives the following upper limits
on $R_V$: 2.2, 2.2$\pm$0.1 and 2.3$\pm$0.9, respectively. All these
values are consistent with the spectrophotometric derivations of $R_V$
for SN 2006X, 2008fp, and 2014J, although we note that the
uncertainties in the Clayton \& Mathis relation are quite large. For
SN~1986G, the cited relations give $R_V$=2.4-2.6, which fully agrees
with the spectrophotometric estimate (Table~\ref{tab:sample}. ; see also
the discussion in Phillips et al. \cite{phillips13}).

The $R_V$ vs. $\lambda_{max}$ relation has been shown to behave
differently in different regions of the Galaxy, to the extent that in
some cases there is a poor (or even no) correlation between the two
quantities (Whittet et al. \cite{whittet01}; Andersson \& Potter
\cite{andersson}). This is interpreted as an indication of a
dependence of the grain alignment on its nature, which brings the dust
composition into the game. The models show that, at least in the
Galaxy, silicates are most likely the only contributors to
polarization (see Voshchinnikov \cite{vosh12} and references
therein). Therefore, while the extinction curve (and hence $R_V$) is
determined by dust grains of all species (silicate, carbonaceous,
iron-rich, etc.) the polarization law is not. Consequently, under
particular conditions, the two can become weakly or even not
correlated (Voshchinnikov \cite{vosh12}). For the sake of
completeness, we mention that none of the cited works include stars
with $\lambda_{max}<$0.39 \mic.

Despite these facts, the four SN we present consistently
show values, of both $\lambda_{max}$ and $R_V$ , that are very uncommon
(or even unprecedented) in our Galaxy. The Galactic sample presented
by Valencic, Clayton \& Gordon (\cite{valencic}) does not include any
object with $R_V\leq$2, while Fitzpatrick \& Massa (\cite{fitz})
report only one such star ($\sim$0.3\% of their sample). This fraction
increases to $\sim$6\% in the anomalous extinction sample by Mazzei \&
Barbaro (\cite{mazzei}). The deviating objects are not associated with
high reddening ($E_{B-V}<$0.5), hence excluding the possibility that
the exceptionally low values derived for our SN are necessarily the
product of large extinction. Although $R_V$ ratios that significantly
deviate from the average are observed in our Galaxy, the distribution
is significantly skewed towards large values. In the Fitzpatrick \&
Massa sample, the fractions of stars with $R_V\geq$3.5 and
$R_V\geq$4.0 are 21.6\% and 9.5\%, respectively. Therefore, the
exceptional behaviours observed for the Type Ia sample are very hard to
interpret in terms of anomalous lines of sight if the host galaxy dust
mixtures were similar to that of the Milky Way.

At face value, these findings lead to the striking, anti-Copernican
conclusion that something is special about the dust of the Milky Way.
Another possibility is that the magnetic field of the Galaxy is special. 
If the field lines are more ordered or the field strength is higher, then the 
alignment of grains and the resulting polarization would be different.
Although not very much is known about extragalactic dust, some
information is available. For instance, Finkelman et
al. (\cite{finkelman}) derived $R_V$=2.8$\pm$0.4 for a sample of nine
early-type galaxies with dust lanes, and concluded that the
extragalactic dust properties are similar to those typical for the
Galaxy. This would suggest that there is something special with the
lines of sight associated with some Type Ia SN. 

The polarization dependencies we report are compatible 
(at least qualitatively) with dichroic polarization by mixtures with an 
enhanced amount of small particles with respect to the typical MW distribution.  
In this respect, to explain the extinction curve of NGC~3034 (M82), 
Hutton et al. (\cite{hutton14}) suggest that  small grains are entrained in
the supernovae-driven wind region and  would reflect the
general properties of the galaxy ISM. However, we emphasize that although the
models with reduced dust sizes can reproduce the required $\lambda_{max}$ shifts
(Siebenmorgen et al. \cite{ralf}, Voshchinnikov \& Hirashita \cite{vosh14}), they
cannot reproduce the increased slope observed in the blue.  
This is particularly true for SN~2014J.

The question as to why the galactic environments of these SN are
systematically different from what is seen in similarly reddened stars
in the Galaxy remains unanswered. Here we only emphasize that the MW
dust properties are deduced from observations of the solar
neighbourhood. No line of sight through any external galaxy will
geometrically ever be similar to any line of sight from the Earth
through the MW.

\subsection{\label{sec:scatt}Alternative explanations}

As shown by Wang (\cite{wang05}), Patat et al. (\cite{patat06b}), and Goobar
(\cite{goobar08}), unusually low values of $R_V$ can be produced by
scattering by circumstellar dust. Although, in principle, this
scenario provides a viable explanation, it leaves a number of open
issues. Most importantly, the propagation of the light echo through the local
dust is expected to produce rapid time variability, both in the
resulting $R_V$ and in the polarization (in case the dust has an
asymmetric geometry). This is particularly true in the early
  phases, where significant variations are expected on the timescales
  of a week (Wang \cite{wang05}; Patat et al. \cite{patat06b}).
Second, in order to have a measurable effect on $R_V$, this requires
conspicuous amounts of dust in the SN vicinity.  This poses problems
both in terms of dust survival and in the implied presence of
significant amounts of gas which, with the possible exception of
SN~2006X, were not detected (see Patat et al. \cite{patat06b} for a
detailed discussion on the subject).  In the cases of 2006X and
2008fp, in any case, it is  difficult to believe that those thick molecular
clouds, where the bulk of reddening and polarization are expected to
arise (Cox \& Patat \cite{cox08,cox14a}), are directly associated with
the SN.  A similar argument applies to the numerous components
detected along the lines of sight to 1986G and 2014J.

The peculiarities of the extinction law along the line of sight to
SN~2014J are discussed by Amannulah et al. (\cite{amanullah}). They
conclude that these are compatible with a power-law
extinction with an index close to $-$2, as expected from multiple
scattering of light (Patat et al. \cite{patat06b}; Goobar
\cite{goobar08}). This conclusion is hard to reconcile with the above
considerations, as it would require that most of the dust responsible
for the observed extinction is confined to distances
$\lesssim$10$^{16}$ cm from the SN. In addition, the large optical
depth implied by the measured extinction ($\tau_{dust}\approx$2) would
translate into a considerable contribution by multiple scattering, which is
an efficient depolarizing mechanism. In these
circumstances, the linear polarization is expected to show a steady
increase from 0.3 \mic\/ to 1.0 \mic\/ (see, for instance, White
\cite{white}; Voshchinnikov \& Karjukin \cite{vosh94}; Kartje
\cite{kartje}; Zubko \& Laor \cite{zubko}), a behaviour that is
observed in reflection nebulae (Zellner \cite{zellner74}). This is
incompatible with the wavelength dependencies discussed here.

Based on multi-wavelength data including UV spectra, Foley et
al. (\cite{foley14}) reached the conclusion that the peculiarities
observed in the extinction curve of SN~2014J can be explained by a
hybrid model. In this scenario, the reddening, which is found to vary
with time, is produced by both circumstellar and interstellar dust,
contributing in roughly similar fractions. While the interstellar
component has rather normal characteristics ($R_V\sim$2.6), the
circumstellar dust has LMC-like properties and is described by a
power-law extinction with an index very close to $-$2, as proposed by
Amanullah et al. (\cite{amanullah}).

In both scenarios, scattering would play a very relevant role and its imprints should
be detectable in the polarization signal. In this respect it is worth discussing the cases
of the two Galactic stars 19 Tau (Matsumura et
  al. \cite{matsumura}) and \#46 in IC~63 (Andersson et
  al. \cite{andersson13}). Both objects are immersed in a reflection
  nebula and their polarization is most likely affected by a
  significant scattering component.  While 19 Tau shows a slow and steady
  linear increase in the polarization from  red to  blue, \#46 in IC~63 displays a very
  steep rise at short wavelengths. The polarization dependence
  of \#46 in IC~63 is well reproduced by a $P(\lambda)\propto \lambda^{-4}$ law in the blue
  range, as expected for both Rayleigh scattering
  on molecules and Mie scattering in the small grain limit (Andersson et al. \cite{andersson13}). 
  This is coupled to a marked wavelength dependence of the polarization angle,
  which changes by $\sim$40 degrees from the B to the U passband
  (Andersson et  al. \cite{andersson13}, their Figure 10). 
  This variation can definitely be excluded in the cases of 
  1986G, 2006X, and 2008fp, while for 2014J this is very unlikely
  (see Figure~\ref{fig:angle}). This indicates that if a scattering component is
  present, it behaves differently from known cases in the Galaxy or its polarization
  angle is, for some reason, parallel to that induced by dichroism.

\begin{figure}
\centering \includegraphics[width=9cm]{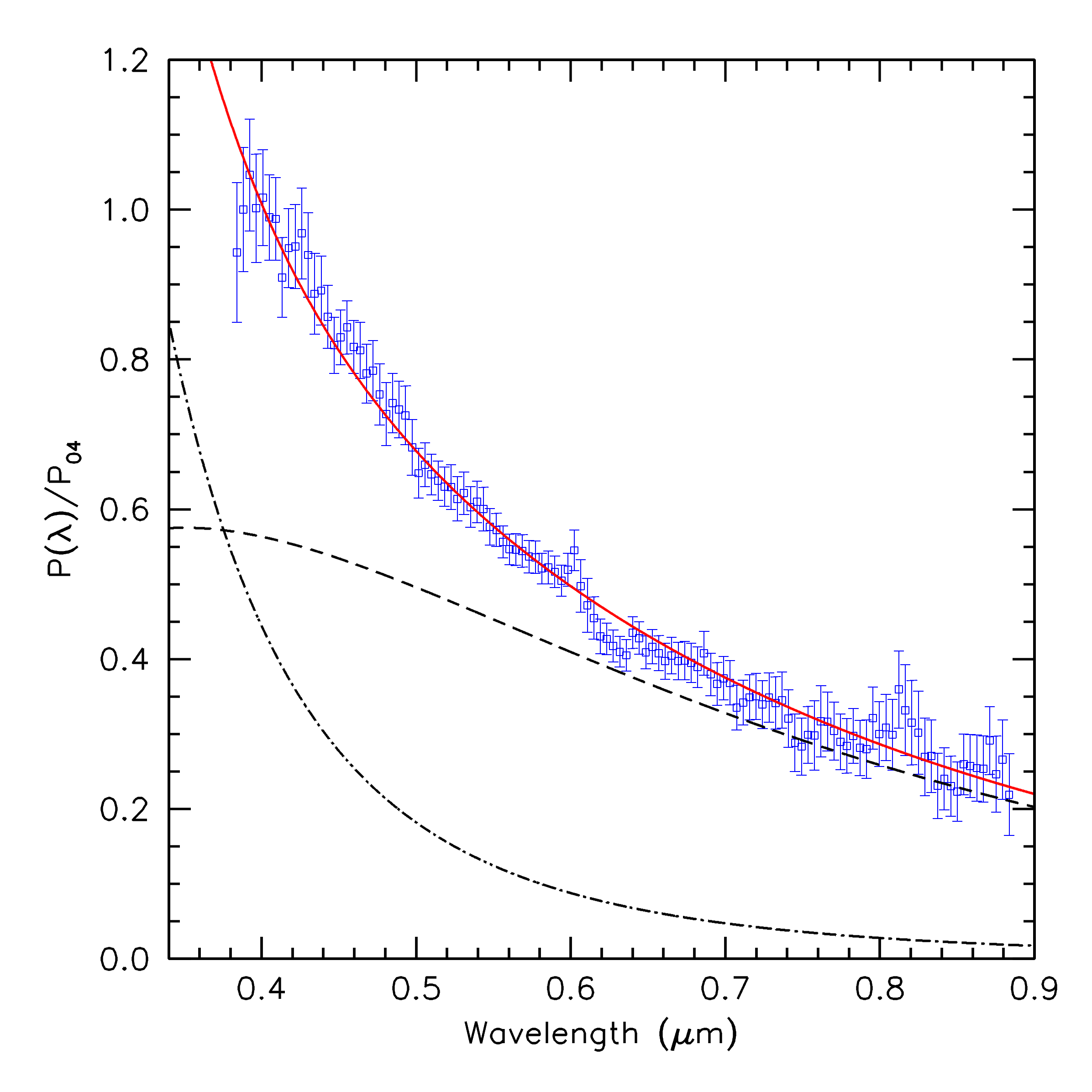}
\caption{\label{fig:scatt} Illustrative decomposition of the observed polarization in SN~2014J using
a Serkowsky law (dashed) and a Rayleigh law (dotted-dashed). The solid line fitting the data traces 
the sum of the two (see text). }
\end{figure}

For illustrative purposes and following Andersson et al. 
(\cite{andersson13}), we applied the decomposition of the observed 
polarization using both a Serkowsky law and a scattering component to the most 
extreme case of SN~2014J. For the scattering we adopted the
following formulation:

\begin{displaymath}
P(\lambda) =  P_s \left (\frac{0.4}{\lambda}  \right)^4
\end{displaymath}

where $\lambda$ is expressed in \mic\/ and we assumed that the polarization 
induced by scattering is parallel to that generated by dichroism. The results are 
shown in Fig.~\ref{fig:scatt}. A reasonable fit can  be reached using the standard 
value for $K$ (1.15),  $\lambda_{max}$=0.35 \mic\/ and $P_s/P_{max}$=0.8. 
Therefore, although a rather blue $\lambda_{max}$ is required, a good match 
can be reached invoking the presence of a scattering component that contributes 
to about 45\% of the polarization at 0.4 \mic. We emphasize that this is a purely
phenomenological decomposition. The wavelength dependency will have to be verified
with detailed modelling, taking the physical properties of dust into account.\ These properties
include the wavelength dependency of the forwards scattering degree that, in turn,
depends on the scattering angle.

Although this finding is intriguing and in line with the scenario proposed by
Foley et al. (\cite{foley14}), there is an aspect that needs to be considered
in addition to those discussed earlier.

In the case in which the dust is mostly placed at very large distances
from the SN, the polarization is acquired purely by transmission
(i.e. through dust dichroism) and the polarization angle reflects the
properties of the galactic magnetic field via the dust grain
alignment.  The picture becomes significantly different if a relevant
fraction of the dust is confined to the immediate surroundings of the
progenitor system. In these circumstances, a non-negligible
contribution to polarization would come from photons scattered into
the line of sight because scattering is a very efficient polarization
mechanism (see Patat \cite{patat05} and references therein).  In that
scenario, the net polarization depends on the geometrical dust
distribution (the polarization is perpendicular to the scattering
plane). On the contrary, the polarization it is null in the case the
geometry is spherically symmetric.  In all other cases, the resulting
net polarization would carry geometrical imprints through the
polarization angle, which for circumstellar material has no relation
to the magnetic field of the host. The fact that the measured position
angles are well aligned with the spiral structure of the hosts places
a stringent constraint on the CSM geometry.

Although the presence of a scattering component in our polarization
data cannot be ruled out\footnote{In this respect, faint interstellar
  light echoes were recently resolved by HST (Crotts
  \cite{crotts14}).}, the scenario proposed by Foley et
al. (\cite{foley14}) for SN~2014J needs to be considered in light of
the results presented here (lack of time variability in the
polarization level and orientation of the electric field).  For
instance, the Foley et al. (\cite{foley14}) scenario could be
reconciled with the polarimetric data if the orientation of the
asymmetric CSM dust responsible for the scattering component is, by
coincidence, not too different from that of the host magnetic
field\footnote{The maximum deviation between two linear polarization
  angles is 90 degrees.}. The polarization angle measured in SN~2014J
is constant to within $\pm$5 degrees across the whole observed
wavelength range (see Figure~\ref{fig:angle}).  This implies that the
two orientations indeed need to be very similar. While we cannot rule
this out for the specific case of 2014J, it seems unlikely that this
is also true for SN~2006X and 2008fp. Based on similar arguments,
Kawabata et al. (\cite{kawabata}) reached a very similar conclusion.
Finally, Brown et al. (\cite{brown14}), on the basis of optical and UV
spectrophotometry, argue that, for the specific case of SN~2014J, most
of the reddening must be produced by interstellar dust.

Future studies of this kind will need to take into account the
important point that dust composition and size distribution are not
fixed in time (Hutton et al. \cite{hutton14}). While large grains
produced by stars initially dominate the size distribution, this is
controlled by ISM processes (Asano et
al. \cite{asano13a,asano13b,asano14}) later on in galaxy history. As a
consequence, the extinction curve is flatter in the earliest stages of
galaxy evolution. The shape of the extinction curve may be related to
the phenomena that lead to dust formation (SN II, AGB stars) and dust
destruction (SN shocks, metal accretion onto dust grains, etc.). Asano
et al. (\cite{asano13b}) find that the dust mass growth in the ISM
becomes effective when the metallicity in a galaxy exceeds a critical
value. This threshold is larger for a shorter star-formation
timescale. Therefore, it is possible that the anomalous behaviour we
report is telling us something about the chemical evolutionary state
of the galaxy, or areas of the galaxy, in which the explosions
occur. We must also take grain growth by accretion and coagulation
(Voshchninnikov \& Hirashita \cite{vosh14}) and alignment mechanisms
(radiation and paramagnetic) on small dust grains (see for instance
Hoang, Lazarian \& Martin \cite{hoang14}) into account.

\section{\label{sec:conc}Conclusions}

In this article we discussed the wavelength dependence of the
insterstellar linear polarization along the lines of sight to four SN
Ia and its implications on the dust properties. The conclusions of
our study can be summarized as follows:

\begin{enumerate}
\item All four SN included in the sample display an anomalous
  interstellar polarization law.
\item While SN~1986G shows a behaviour that is compatible with the most
  extreme cases known in the Galaxy, SN 2006X, 2008fp, and 2014J
  display an unprecedented wavelength dependence, with $\lambda_{max}<$0.4 \mic.
\item The total-to-selective absorption ratio $R_V$ derived from 
    spectrophotometry is exceptionally low for all four events.
\item In SN 2006X and 2008fp, the bulk of the polarization most likely arises
within thick molecular clouds. The anomalous polarimetric behaviour may be
associated with the peculiar properties of these clouds.
\item In SN 1986G and 2014J, the polarization is more probably
  generated within multiple components of similar optical depth along
  the line of sight.  For these events, it is hard to argue in favour of
  peculiar properties of a single cloud. More plausibly, the
  observations reflect global characteristics of the hosts.
\item The very small size of the sample does not allow us to establish a
  relation between the anomalous polarimetric properties and
  the morphological type of the hosts.
\item Although the objects in the sample are significantly reddened
  (0.6$\leq E_{B-V}\leq$1.4), this does not explain the observed
  behaviour in terms of systematic effects related to high
  dust optical-depth lines of sight.
\item The dust responsible for the anomalous polarization properties is most
  probably of interstellar nature. A similar conclusion was reached by Kawabata
  et al. (\cite{kawabata}).
\item Although the presence of circumstellar dust cannot be excluded, the
lack of time evolution and the alignment with the local spiral arm pattern pose
rather stringent constraints on its geometrical distribution.
\item The extreme case of SN~2014J can be explained, at least
  phenomenologically, invoking a significant contribution by
  scattering, provided that an ad hoc distribution for the local dust
  is adopted.
\end{enumerate}

In general, the observed polarization properties (at least for SNe
1986G, 2006X and 2008fp) can be understood in terms of an enhanced
abundance of small grains with respect to the typical MW dust
mixture. However, dust size alone cannot explain the observed
wavelength dependence. The reasons why all SN in the sample show a
behaviour so different from that seen in the Galaxy are unclear. The
chemical/physical evolution of the hosts may be playing a relevant
role. Future studies will have to proceed on two different fronts: a)
increase the sample of well -studied Type Ia SN to investigate in a
statistically robust way possible correlations between dust
properties, galaxy types, and SN characteristics; b) obtain
high-quality spectropolarimetry of low $R_V$ Galactic objects to
establish possible connections to what is seen in extragalactic
environments.

\begin{acknowledgements}
This paper is dedicated to the memory of Prof. Guido Barbaro, who
first introduced F.~P. to the physics of the interstellar
medium. Mandi, Guido. The authors are grateful to Dr G-B Andersson for
illuminating discussions.

The German-Spanish Astronomical Center, Calar Alto (Spain) is jointly
operated by the Max-Planck-Institut f\"ur Astronomie Heidelberg and
the Instituto de Astrofisica de Andalucia (CSIC). The Mercator
Telescope is operated on the island of La Palma by the Flemish
Community, at the Spanish Observatorio del Roque de los Muchachos of
the Instituto de Astrofisica de Canarias. The HERMES spectrograph is
supported by the Fund for Scientific Research of Flanders (FWO),
Belgium, the Research Council of KU~Leuven, Belgium, the Fonds
National de la Recherche Scientifique (F.R.S.-FNRS), Belgium, the
Royal Observatory of Belgium, the Observatoire de Gen\'eve,
Switzerland, and the Th\"uringer Landessternwarte Tautenburg,
Germany. The authors are grateful to the Calar Alto Observatory and
its staff for carrying out the observations in a very efficient way.

S.~T. acknowledges support by the Trans-regional Collaborative
Research Centre TRR33 of the German Research Foundation (DFG). The
research of J.~C.~W. is supported in part by NSF Grants AST-1109801
and No. PHYS-1066293, and the hospitality of the Aspen Center for
Physics extended to J.~C.~W.

\end{acknowledgements}

\end{document}